# Multiplexed Five-Color Molecular Imaging of Cancer Cells and Tumor Tissues with Carbon Nanotube Raman Tags in the Near-Infrared

Zhuang Liu,[1,4] Scott Tabakman,[2,4] Sarah Sherlock,[2] Xiaolin Li,[2] Zhuo Chen,[2] Kaili Jiang,[3] Shoushan Fan,[3] Hongjie Dai[2*]

1, Functional Nano & Soft Materials Laboratory (FUNSOM), Jiangsu Key Laboratory for Carbon-Based Functional Materials & Devices , Soochow University, Suzhou, Jiangsu, China

2, Department of Chemistry, Stanford University, Stanford, California, USA

3, Department of Physics and Tsinghua-Foxconn Nanotechnology Research Center, Tsinghua University, Beijing, China

4, These authors contributed equally to the work

* Correspondence should be sent to: hdai@stanford.edu

**Abstract**

**Single walled carbon nanotubes (SWNTs) with five different C13/C12 isotope compositions and well-separated Raman peaks are synthesized and conjugated to five targeting ligands to impart molecular specificity. Multiplexed Raman imaging of live cells is**

**carried out by highly specific staining of cells with a five-color mixture of SWNTs.** *Ex vivo* **multiplexed Raman imaging of tumor samples uncovers a surprising up-regulation of epidermal growth factor receptor (EGFR) on LS174T colon cancer cells from cell culture to** *in vivo* **tumor growth. This is the first time five-color multiplexed molecular imaging is performed in the near-infrared (NIR) region under a single laser excitation. Near zero interfering background of imaging is achieved due to the sharp Raman peaks unique to nanotubes over the low, smooth auto-fluorescence background of biological species.**

## Introduction

Molecular imaging by optical techniques plays important roles in biology and medicine owing to high spatial resolution and detection sensitivity, providing not only geometrical but also functional information of biological and medical systems [1, 2]. However, various challenges remain in this area. Photo-bleaching is a problem for most organic fluorescent dyes, preventing long-term imaging and monitoring of fluorescently labeled substances [3]. High background due to auto-fluorescence of biological tissues also presents a problem to fluorescence imaging [4]. For *in vivo* imaging, the depth of tissue penetration by light is limited. The near-infrared (NIR) region from 700 nm to 1400 nm is an ideal window for optical imaging of biological systems due to the minimal tissue absorption and low auto-fluorescence in this region [5].

Multicolor multiplexed molecular imaging is desired to simultaneously map out different species and reveal the geometrical and functional relations between different targets. Ideally, multicolor imaging of various molecular targets under a single excitation with both excitation and emissions in the narrow NIR range is desired. Although multi-color fluorescence imaging has been achieved using different fluorescent dyes or quantum dots with excitations/emissions in the visible to NIR range (400-900 nm) [6-8], imaging multiplicity is limited in the narrow NIR window due to spectral overlay between fluorescence peaks with full-width half-maximum (FWHM) as large as 50-100 nm. Also, auto-fluorescence, although low in the NIR region, still exists and varies between species in a biological sample [4]. Complex algorithms are needed to differentiate the background for fluorescence-based imaging [7].

Raman scattering emissions have narrow spectral features with FWHM within a few nm, allowing for increased imaging multiplicity [9]. Six or even more Raman 'colors' have been obtained by conjugating different Raman active dyes on Ag or Au nanoparticles and simultaneously imaged on the substrate [10]. Three-color non-specific Raman imaging under the skin of animals with NIR excitation and emissions has also been shown [11]. Thus far, multiplexed, multi-color (>3) targeted molecular Raman imaging of biological systems with high selectivity has not been achieved.

In recent years, carbon nanotubes have been explored in biological applications including molecular delivery [12-16], sensing and detection [17, 18], and imaging [19, 20]. SWNTs are one dimensional materials [21] with unique intrinsic optical properties such as photoluminescence in the NIR [19, 22] and strong resonant Raman signatures [23], both of which are useful for biological detection and imaging. We have shown that SWNTs with three different C13/C12 isotope compositions exhibit shifted Raman G-band peaks, and can be used for three-color multiplexed Raman imaging of live cells [24].

In this work, five SWNT samples with various C13/C12 ratios were synthesized, functionalized, and conjugated with different targeting ligands for specifically labeling and imaging of biological samples. Relative protein expression levels on several human cancer cell lines were analyzed. Five-color multiplexed Raman imaging was carried out for cancer cells in culture and in tumors *ex vivo*. We observed tumor vasculatures labeled by one SWNT color and cancer cells by other colors. Unexpectedly, we observed a drastic up-regulation of EGFR/Her1 expression on LS174T human colon cancer cells during tumor growth in nude mice (not present *in vitro* in cell culture). This is the first multi-color molecular Raman imaging of five different biological targets with excitation/emission in the NIR region under a single laser excitation, with essentially zero interfering background.

## 1. Results and Discussion

### 1.1. Five unique, target specific SWNT Raman scattering labels

Isotopically modified SWNTs were synthesized by using a mixture of C13 methane and C12 methane in chemical vapor deposition (CVD) growth of nanotubes [24, 25]. SWNT samples grown with gas phase C13 ratios of 100%, 65%, 50%, 25% and 0% exhibited Raman G-band peaks at 1529 cm$^{-1}$, 1546 cm$^{-1}$, 1559 cm$^{-1}$, 1575 cm$^{-1}$ and 1590 cm$^{-1}$, which were defined as Color 1, Color 2, Color 3, Color 4 and Color 5, respectively (Fig. 1). As grown SWNTs were purified as reported earlier [25]. Phospholipid-polyethylene glycol-amine (PL-PEG5000-Amine) was used to functionalize SWNTs and afford water soluble SWNTs (length 50~300 nm) after centrifugation to remove large bundles and filtration to remove excess surfactant [24, 26-28]. The five functionalized SWNT samples were conjugated to five different targeting antibodies and a peptide, including Erbitux (anti-EGFR/Her1) to target epidermal growth factor receptor (EGFR/Her1), cyclic arginine-glycine-aspartic acid (RGD, cyclo-RGDyK) peptide to target integrins such as $\alpha_v\beta_3$, goat anti-human CEA antibody to recognize carcinoembryonic antigen (CEA), Rituxan (anti-CD20) to target CD20 protein and Herceptin (anti-Her2) to target Her2/neu protein (Fig. 1a) [19, 24].

## 1.2. Five-plexed *in vitro* Raman imaging and semi-quantitative analysis of cancer cell surface biomarkers

We carried out multiplexed two-color cell imaging of Her1 and Her2, two of the Her-family antigens on various cancer cells. Six different human cancer cell lines were incubated with a mixture of Color1-anti-EGFR/Her1 and Color5-anti-Her2 at 4 $^o$C for 1 hour and then washed by PBS. Spectroscopic mapping of cells over a large area was then carried out. The smooth auto-fluorescence background from 1470 cm$^{-1}$ to 1640 cm$^{-1}$ was subtracted to reveal pure SWNT G-band Raman peaks (see Electronic Supplementary Material Fig. S1, S5). The G peak intensities of C13 and C12 SWNTs at each location were then integrated and used to construct two-color Raman images (Fig. 2a). Strong Color1 signals (assigned to red) but not Color5 (green) were observed from BT20 and MDA-MB-468 cells, suggesting high Her1, but low Her2 expressions on these cells (Fig. 2a, 2$^{nd}$ and 6$^{th}$ rows). MDA-MB-231 cells also showed moderate expression of EGFR/Her1 (Fig. 2a, 3$^{rd}$ row). On the other hand, high Her2 expression was noted on BT474 and MDA-MB-361 cells as evidenced by the observed strong Color 5 (Her2) signals on those cells (Fig. 2a, 1$^{st}$ and 4$^{th}$ rows). Other cell lines exhibited low Her2 expression (Fig. 2a).

Since the SWNT concentrations used in the staining solution were far beyond the saturation levels under our experimental conditions (Electronic Supplementary Material Fig S3), each cell surface receptor should have bound an antibody-conjugated nanotube, and because the Raman intensity of SWNTs is linearly dependent on the SWNT concentration, the averaged Raman intensity per cell was proportional to the mean receptor level on cells, which allowed us to quantify relative protein expression levels of different cell lines. The EGFR/Her1 expression levels followed a trend of MDA-MB-468 > BT20 > MDA-MB-231 > MDA-MB-435 > BT474 ~ MDA-MB-361. In the case of Her2, we observed BT474 > MDA-MB-361 > BT20 ~ MDA-MB-231 ~ MDA-MB-468 ~ MDA-MB-435 (Fig. 2b). Those trends were consistent with flow cytometry data with fluorescently labeled antibodies (Electronic Supplementary Information, Fig. S4), as well as previous reports [29], suggesting that semi-quantitative measurement of surface protein expression levels can be achieved by SWNT based labeling and Raman spectroscopic imaging.

Next, we used five-plexed SWNT Raman tags to label various cancer cells over-expressing different receptors. After incubation with the five-color SWNT mixture (Color1-anti-EGFR/Her1, Color2-RGD, Color3-anti-CEA, Color4-anti-CD20, Color5-anti-Her2), cells were washed by PBS and subjected to Raman spectroscopic imaging. The Raman spectrum at each pixel was de-convoluted to those of the five Raman colors after background subtraction (see method section for detailed data processing), affording five-color Raman images, with relative Raman intensities of de-convoluted spectra at each pixel (Fig. 3). The five cancer cell lines with different surface protein expression profiles were successfully labeled by SWNT tags with high specificity, showing distinct Raman 'colors' on MDA-MB-468, U87MG, LS174T, Raji and BT474 cells, respectively (Fig. 3, top 5 rows). Minimal crosstalk between different colors was seen by analysis of the spectra. The non-specific binding (NSB) of mis-matched nanotube conjugates on cells was negligible (positive binding over NSB signals = 50~100:1). When the five cell lines were mixed together and stained by a mixture of five-color SWNT conjugates, Raman spectroscopic imaging clearly differentiated five different types of cancer cells by their respective SWNT Raman labels with minimal overlapping of colors between cell types. Five-color multiplexed molecular Raman imaging of cells *in vitro* is thus achieved by SWNT Raman labels.

## 1.3. Molecular imaging of e*x vivo* tumor slices via five-plexed SWNT Raman labels

To explore the utility of multi-color SWNT Raman tags for probing samples with high biological and medical relevance, we used the five-plexed SWNTs for *ex vivo* tumor staining and Raman imaging, aimed at detecting multiple species related to cancer in a single experiment, and resolving the locations and spatial relations of various molecular targets in the tumor. Nude mice bearing LS174T human color cancer tumors were sacrificed 15 days after tumor inoculation. The collected tumors were frozen-sliced into 5 μm thick tumor slices and stained with the five-color, five-plexed SWNT mixture (see Method). Raman spectroscopic mapping of the slices was then preformed at both high resolution over a small area and low resolution over the entire slice (Fig.4b and c, respectively), yielding multi-color Raman images of protein expression. As expected from the *in vitro* cell imaging results (Fig. 3), cancer cells in LS174T tumor slices showed a high level of CEA expression (Fig. 4, blue color). Tumor vessel-like structures were positively stained by Color5-RGD conjugates, indicating high integrin expression on the tumor vasculature, most likely the integrin $\alpha_v\beta_3$ [30, 31] (Fig. 4, red color). It is well known that integrin $\alpha_v\beta_3$ receptor regulates tumor angiogenesis and is over-expressed on various tumor vasculature endothelial cells *in vivo* [32, 33].

Interestingly, we found that EGFR/Her1 surface receptor, which had low expression *in vitro* on LS174T cells [34, 35], was highly up-regulated *in vivo* in the xenograft tumor probed 15 days post inoculation, and was clearly stained by Color1-anti EGFR/Her1 (Fig. 4). The high degree of overlay of Color1 (yellow, EGFR/Her1) and Color3 (blue, CEA) in the Raman images (Fig. 4b&c) suggested co-localization of EGFR/Her1 and CEA over-expression on LS174T cancer cells, with the former developed *in vivo*. In addition, LS174T tumor slice was also diffusely stained by Color2-anti-Her2, suggesting a moderate up-regulation of Her2 receptor in the xenograft LS174T tumor model. In the literature, only one recent work reported high tumor uptake of anti-EGFR/Her1 antibody in the LS174T xenograft mouse model through an *in vivo* biodistribution study, despite the low EGFR/Her1 expression level on LS174 cells *in vitro* [35]. Similar observations have been made for low levels of Her2 in the same tumor model [36]. However, no prior imaging result exists to show the spatial relations of CEA and EGFR expression in this xenograft tumor model. Our finding suggests that the expression profile of cancer cell receptors could change from cell culture to tumor growth *in vivo*. The over-expressions of EGFR and Her2, which play important roles in the process of cancer cell proliferation, is associated with many types of cancers and has become one of the most important targets in cancer therapy [37]. Our observation suggests the importance of *in vivo* and *ex vivo* probing and evaluation of EGFR for both diagnostic and targeted treatment proposes. It also highlights the usefulness of multiplexed imaging to probes multiple targets with high efficiency and spatial resolutions to glean valuable insights into the complexities of molecular pathways and networks in tumors.

## 2. Conclusion

Five-color multiplexed imaging in the narrow NIR window has not been achieved by fluorescence imaging using either organic dyes or quantum dots owing to the wide fluorescence spectra relative to the narrow NIR window. In our Raman imaging, excitation (785 nm) and scattered photons (892 nm-897 nm) are all in the narrow NIR window with low tissue absorption and auto-fluorescence background (Electronic Supplementary Information, Fig S5). SWNTs have a single, narrow Raman G-band peak with FWHM < 2 nm, allowing for high degrees of multiplicity. As many as five Raman colors are simultaneously imaged under a single laser excitation for measuring five different targets in a single Raman mapping scan. Although previous proof-of-concept work have shown that Raman tags based on molecules and metal nanoparticle complexes can be used for multiplexed imaging with many colors [10, 11], our work is the first to use multiple (>3) Raman colors successfully labeling different biological species with specificity to afford multiplexed molecular imaging. Nanotube Raman tags are advantageous over common Raman dye molecules, such as cyanine dyes and rhodamine derivatives, as they present simple yet intense spectra, without the complex, multiple Raman peaks and high fluorescence emission background of the latter. The simple NIR Raman scattering spectra of SWNTs allow for facile spectral deconvolution without the need of sophisticated algorithms, including those necessary to map out sample auto-fluorescence, which varies in space over biological samples [4, 7]. Also important is that nanotube Raman signals exhibit no quenching or bleaching under a wide range of imaging conditions [38].

The sharp single G-peak feature in the SWNT Raman spectrum enables us to easily distinguish SWNT Raman signals from the low and slow varying auto-fluorescence background, allowing for 'background free' optical imaging with high imaging and detection sensitivities (Electronic Supplementary Material Fig. S1). On the other hand, the Raman spectral acquisition times reported in this manuscript are substantially longer than those used in confocal fluorescence microscopy, which represents a practical limitation of the approach. Promising non-linear Raman measurement techniques [39] have recently been reported which may potentially reduce acquisition times and improve the utility of multiplexed SWNT Raman labels.

Besides utilizing the isotope-dependent Raman G peak shift, the SWNT radial breathing mode (RBM) peak position exhibits a systematic diameter dependence [23]. Even more Raman colors could be achieved for chirality/diameter separated SWNTs [40]. Taken together, SWNTs are promising high performance NIR Raman tags for multi-color multiplexed optical imaging and detection *in vitro*, *ex vivo* and potentially *in vivo*.

## 3. Methods

### 3.1. Synthesis of isotopically modified SWNTs

The synthesis of isotopically modified SWNTs was preformed on a silica support Fe-Ru bimetallic catalyst using $CH_4$ chemical vapor deposition (CVD) growth following our previously reported protocol at 850 $^oC$ [25]. To control isotope compositions of SWNTs, C12 $CH_4$ gas and C13 $CH_4$ gas (99% pure) were mixed at desired ratios by flow controllers during SWNT growth. The as-grown product was etched in 15% HF solution in water (75%) and ethanol (10%) overnight at room temperature to remove the silica support. The purified SWNTs were then washed thoroughly with ethanol and water to neutral pH and stored for further use. Note that pure C12 Fe-Ru SWNTs were used in cell incubation and imaging (Fig. 2 & 3) while Hipco SWNTs (pure C12) from Carbon Nanotechnology Inc were used as color 5 (Color5-RGD) in tumor slice staining and imaging (Fig. 4).

### 3.2. Functionalization and bioconjugation

Functionalization and bioconjugation of SWNTs have been well established in our previous reports [14, 19, 27, 28]. DSPE-PEG5000-$NH_2$ (NOF cooperation) at a concentration of 1 mg/ml was sonicated with SWNTs from above in water for 1 h. Water bath was changed every 20 minutes to avoid over-heating. The nanotube suspension was then centrifuged at 24,000g for 6 h to remove aggregates and metal catalysts and collect the supernatant. Excess DSPE-PEG5000-$NH_2$ was removed by filtration through 100 kDa MWCO ultracentifugal filter devices (Millipore Amicon Ultra) prior to conjugation to antibody or peptide.

Erbitux (anti-Her1), Rituxan (anti-CD20) and Herceptin (anti-Her2) were gifts from Drs. Dean Felsher and Sanjiv Gambhir (Stanford University Medical Center, Stanford, CA, USA). Polyclonal goat anti-CEA antibody was purchased from Biospacific Inc (Emeryville, CA, USA). These antibodies are thiolated by mixing them with 2-Iminothiolane hydrochloride (Traut's Reagent, Pierce) at 1:10 molar ratio (all IgG masses assumed to be 150 kDa) in PBS in the presence of 20 mM EDTA for 2 h. Un-reacted Traut's Reagent was removed by filtration through a 100 kDa MWCO ultracentrifugal filter devices. The thiolated antibodies were used immediately in the following conjugation. Thiolated RGD (RGD-SH) was a gift from Dr. Xiaoyuan Chen, (NIBIB/NIH, Bethesda, MD, USA) and used directly without further modification.

Functionalized SWNTs at concentrations of 300 nM (calculated by extinction at 808 nm [37], $\varepsilon_{808nm} = 0.0079$ $nM^{-1}$ $cm^{-1}$) were reacted with 1 mM sulfo-succinimidyl 4-N-maleimidomethyl cyclohexane-1-carboxylate (Sulfo-SMCC, Pierce) for 2 h in PBS at pH 7.4. After thorough removal of excess sulfo-SMCC by filtration through 100 kDa MWCO ultracentrifugal filters devices, the SWNT solutions were mixed with thiolated antibodies in PBS in the presence of 5 mM EDTA. In the reaction buffer, the final SWNT concentration was 300 nM, while that of antibodies was 1.5 μM (~1:5 molar ratio). The reactions were allowed to proceed overnight at 4 °C, affording SWNT-antibody conjugates used for cell incubation and imaging. RGD-SH was conjugated to SWNTs using a similar protocol except that 100 μM of RGD-SH was used in the reaction mixture. 5 mM tris(2-carboxyethyl) phosphine hydrochloride (TCEP) was added to prevent the oxidation of the thiols.

### 3.3. Cell culture and incubation

BT20, BT474, MDA-MB-231, MDA-MB-361, and MDA-MB-468 human breast cancer cell lines, MDA-MB-435 M14 melanoma cell line, U87MG brain cancer cell line, LS174T human colon cancer cell line and Raji human B-cell lymphoma cell line were either obtained from the American Type Culture Collection (ATCC) or gifts from Drs. Xiaoyuan Chen (NIBIB/NIH, Bethesda, MD, UDA) and Dean Felsher (Stanford University Medical Center, Stanford, CA, USA). Descriptions of cell propagation and subculturing conditions may be found in the Electronic Supplementary Material.

For cell incubation, 200 μl of cells (~1 million/ml) were incubated with 50 μl of SWNT solution (containing either a two-color or five-color SWNT mixture) in PBS for 1 hour at 4$^o$C. The final concentration of each color of SWNT (average length ~150 nm) [41] in the incubation solution was 10 nM. Cells were washed with PBS three times before Raman imaging.

### 3.4. Raman imaging

Confocal Raman spectroscopic imaging was carried out using a Horiba-Jobin-Yvon Raman confocal microscope with a 785 nm laser (80mW) as the excitation light source. A drop of cell suspension was sealed between two thin plastic cover-slides for imaging. A 50× objective was used with ~1 μm laser spot size. A 1 mm pin-pole was applied to afford a spatial resolution of ~1 μm in xyz-axis. Each Raman spectroscopic map contains at least 100 × 100 spectra with 0.5 s integration time for each spectrum. 2 μm and 1 μm step sizes were used for Raman mapping to obtain large area images (Fig. 2, last row of Fig. 3, and Fig. 4b) and zoomed-in images (Fig. 3, top five rows), respectively. 40 μm steps were used for whole tissue slice mapping (Fig. 4c).

### 3.5. Background subtraction

The auto-fluorescence background in the Raman spectrum of cell or tissue images was low and smooth in the range of interest from 1470 cm$^{-1}$ to 1640 cm$^{-1}$, which is the Raman G-band region of SWNTs. We simply subtracted a line from 1470 cm$^{-1}$ to 1640 cm$^{-1}$ as to remove the smooth background at each pixel (Fig. S1), producing essentially background free spectra used for Raman image construction and data analysis. A typical spectrum of positively stained cells by SWNTs exhibits 1000~5000 G-band Raman counts, while the standard deviation of background noise in the spectrum of non-treated cells is 20~40. The signal to noise ratio is ~100 at each pixel.

### 3.6. Data analysis for two color Raman imaging

Raman spectra of Color-1 (pure C13) and Color-5 (pure C12) have little overlay between each other. Intensities of individual peaks (1510-1550 cm$^{-1}$ for Color1, 1570-1610 cm$^{-1}$ for Color5) were integrated independently at each pixel for constructing two-color images shown in Fig. 2. Red and green colors were artificially assigned to Color1 and Color2 SWNT signals, respectively.

For semi-quantitative analysis of relative protein expression levels, all spectra in each spectroscopic map (Fig. 2a) were summed. The G-band peak height in the sum spectrum was determined as the total Raman counts for the whole map. The total number of cells in the mapping area was counted from the corresponding optical bright field image, and used to calculate an averaged Raman counts per cell, which was used to determine the relative protein level on cells after normalization of SWNT Raman scattering intensity factor (Electronic Supplementary Material Fig S2). Raman scattering intensity factors of the various SWNT materials were determined empirically by normalizing suspension absorbances at the excitation wavelength (785 nm), and were linearly proportional to SWNT concentration, determined by UV-Vis-NIR absorbance.

### 3.7. Data analysis for five-color Raman imaging

To create a five-color image based upon relative scattering intensities, the spectra of five pure SWNT solutions (after background subtraction) at the same OD at 785 nm (Electronic Supplementary Material Fig. S2) were used for deconvolution of a background-subtracted, recorded spectrum, to obtain five Raman G peak components at each mapping site, thus accounting for the Raman scattering intensity factor of each SWNT material. The software generated relative Raman intensity values (the deconvoluted intensity divided by the input spectral intensity) of five colors at each pixel. A color scale was then assigned to each of the five peaks corresponding to the "percentage of the maximum" deconvoluted intensity of the entire image (100% represents the intensity of the strongest pixel among all 5 colors in an image). Separate and overlaid five-color Raman images were then constructed.

**3.8. Tumor inoculation, slicing and staining**

Athymic nude mice obtained from Harlan Sprague Dawley Inc. were housed in Stanford Research Animal Facility (RAF) under Stanford Institutional Animal Care and Use Committee (IACUC) protocols. Approximately $5\times10^6$ LS174T cells were implanted subcutaneously on the shoulder of nude mice. Mice were sacrificed 15 days after tumor inoculation. Tumors taken from the mice were immediately placed into optimal cutting temperature (OCT) medium and frozen by dry ice. 5 µm thick tumor slices were cut by a microtome, placed on quartz slides, and fixed with dry ice-cold acetone. After blocking with 3% fetal bovine serum (FBS) and 0.1% Tween-20 in PBS for 30 min at room temperature, tissue slides were incubated with a solution of 5-color SWNT mixture in RMPI-1640 cell medium with 10% FBS for 1 hour. The total SWNT concentration in the incubation solution was 20 nM (4 nM for each color of SWNT conjugate). These tissue slides were then washed with PBS for 3 times (5 minutes each time) and briefly with distilled water. Slides were dried in air for 30 minutes before Raman spectroscopic imaging.

## Acknowledgement

This work was supported partially by CCNE-TR at Stanford University, NIH-NCI R01 CA135109-02 and Ensysce Biosciences Inc.

**Figure Captions:**

**Figure 1.** SWNTs with different Raman 'colors'. **(a)** Scheme of isotopically modified SWNTs, grown from FeRu catalysts, conjugated with different targeting ligands. Color1, 2, 3, 4 and 5 represented FeRu-grown SWNTs with C13 percentages (determined by the C13-methane / C12-methane growth gas ratios) of 100%, 65%, 50%, 25% and 0%, respectively. **(b)** Raman spectra of the five different SWNT samples in aqueous solutions. The shift of SWNT Raman G-band peak was clearly dependent on the C13/C12 ratio in SWNTs. The average shift between two adjacent colors was about 15 $cm^{-1}$.

**Figure 2.** Two-color Raman imaging of Her1 and Her2 of different breast cancer cell lines. **(a)** Two-color Raman images of six breast cancer cell lines. Cells were stained by a mixture of Color1-anti-EGFR/Her1 and Color5-anti-Her2 conjugates for Raman spectroscopic imaging. 150-400 cells were imaged in each mapping field. Scale bar = 50 μm. **(b)** Semi-quantitative analysis of two-color Raman imaging data. Averaged Raman intensity per cell of each breast cancer cell line was calculated by dividing the sum of Raman intensity in each spectroscopic image by the total number of cells in the mapping area counted from the related optical image (see method section), and normalized by Raman scattering intensity factors of two types of SWNTs (Electronic Supplementary Material Fig. S2). Relative EGFR/Her1 and Her2 expression levels were determined by the average Raman intensities of Color1 and Color5 per cell, respectively. Error bars were based on triplicated images.

**Figure 3.** Five-color Raman imaging of cancer cells. In the top 5 rows (scale bars = 10 μm), each cell line was stained by a 5-color multiplexed SWNT mixture for Raman spectroscopic imaging. In the bottom row (scale bars = 40 μm), 5 types of cancer cells were mixed together and the incubated with the 5-color SWNT mixture. Spectral deconvolution was used to resolve different colors of SWNTs (see Electronic Supplementary Material). The colors are scaled as the percentage of the maximum relative Raman intensity of the deconvoluted spectra (most intense spectrum of any color is assigned 100%). Spectra of five SWNT solutions at the same OD the excitation wavelength were inputted to deconvolute the raw data (Electronic Supplementary Material Fig S2.)

**Figure 4.** Multi-color *ex vivo* tumor imaging by Raman. Note that SWNTs conjugates used here were different from those used in the previous cell imaging experiments. Hipco SWNTs (C12) with a greater Raman scattering intensity factor were used as color 5. **(a)** A photo of a LS174T tumor bearing nude mouse used in this study. The arrow is pointed to a tumor. **(b)** Five-color Raman images of a LS174T human colon tumor slice from a mouse model. Separate images representing five protein expression levels were generated by deconvoluting the raw spectra at each pixel into fractional percentages of the five intensity factor-corrected, G band spectra. Colors are plotted as percentage of maximum intensity (most intense pixel of any color is assigned 100%). The overlay image exemplifies the high degree of correlation between CEA

expression and EGFR expression. Note that in overlay images Color5-RGD intensity is doubled for clarity. LS174T cells in the tumor showed high levels of EGFR/Her1 (yellow) and CEA (blue) receptors, and co-localization. Integrin expression visualized by Color5-RGD Raman staining (red) appeared to be on the tumor vessels. A relatively low level of Her2 expression was also observed on LS174T tumor cells. Right: a corresponding bright field optical image of the tumor slice. The highlighted area was a tumor vessel. Integrin expression, likely that of $\alpha_v\beta_3$, was clearly associated with blood vessels in the tumor. Scale bar = 50 μm **(c)** Large area whole tumor slice five-color Raman images. Five separate images for each protein expression level were generated as in Fig 4b, and are accompanied by an overlay image, in which Color5-RGD intensity has been doubled for clarity. The bright field image of the whole tumor slice was obtained by stacking multiple small area images taken under a 10 x objective. Blood vessels going through the tumor were clearly visualized in Color5 image. Scale bar = 1 mm.

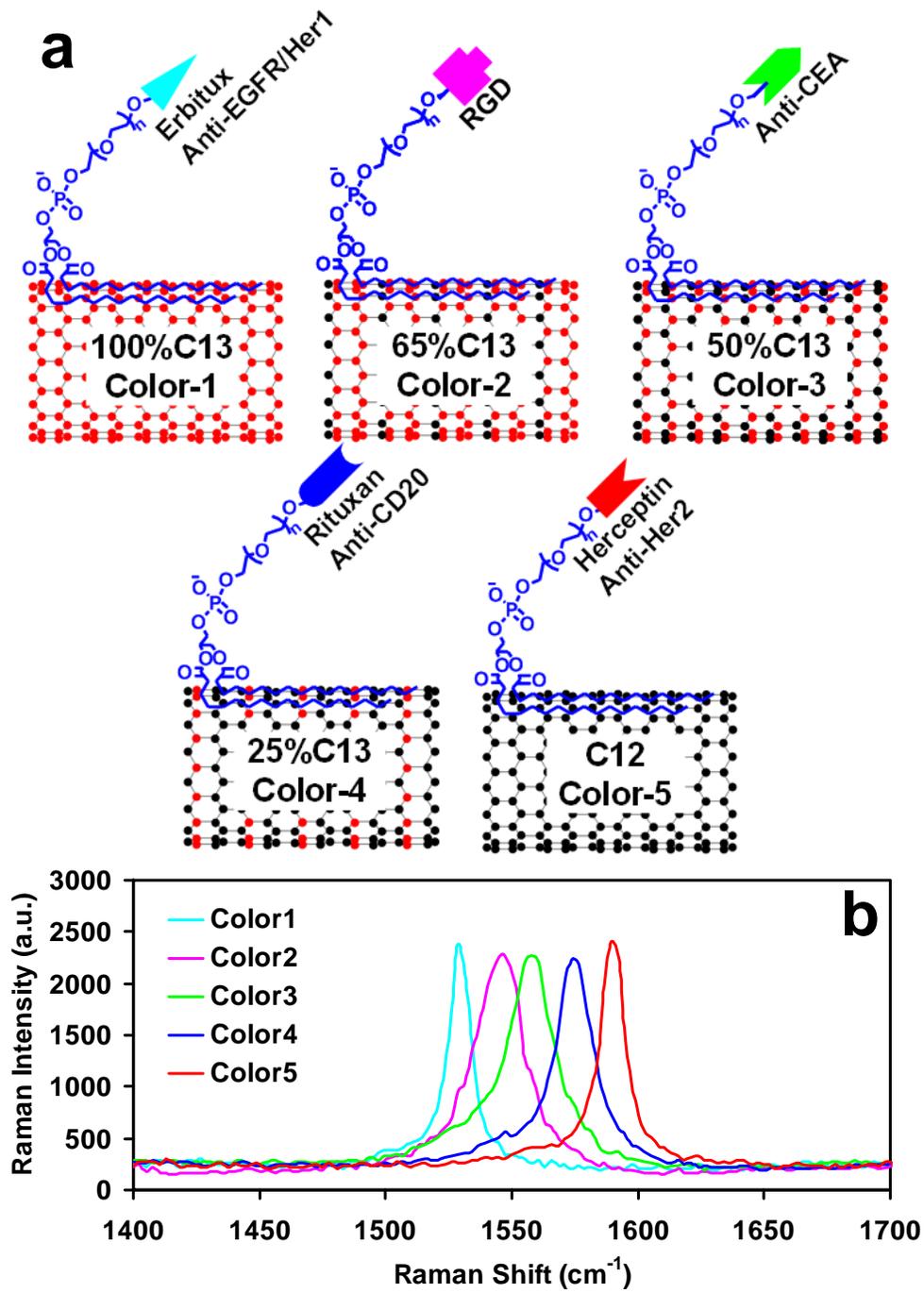

Figure 1

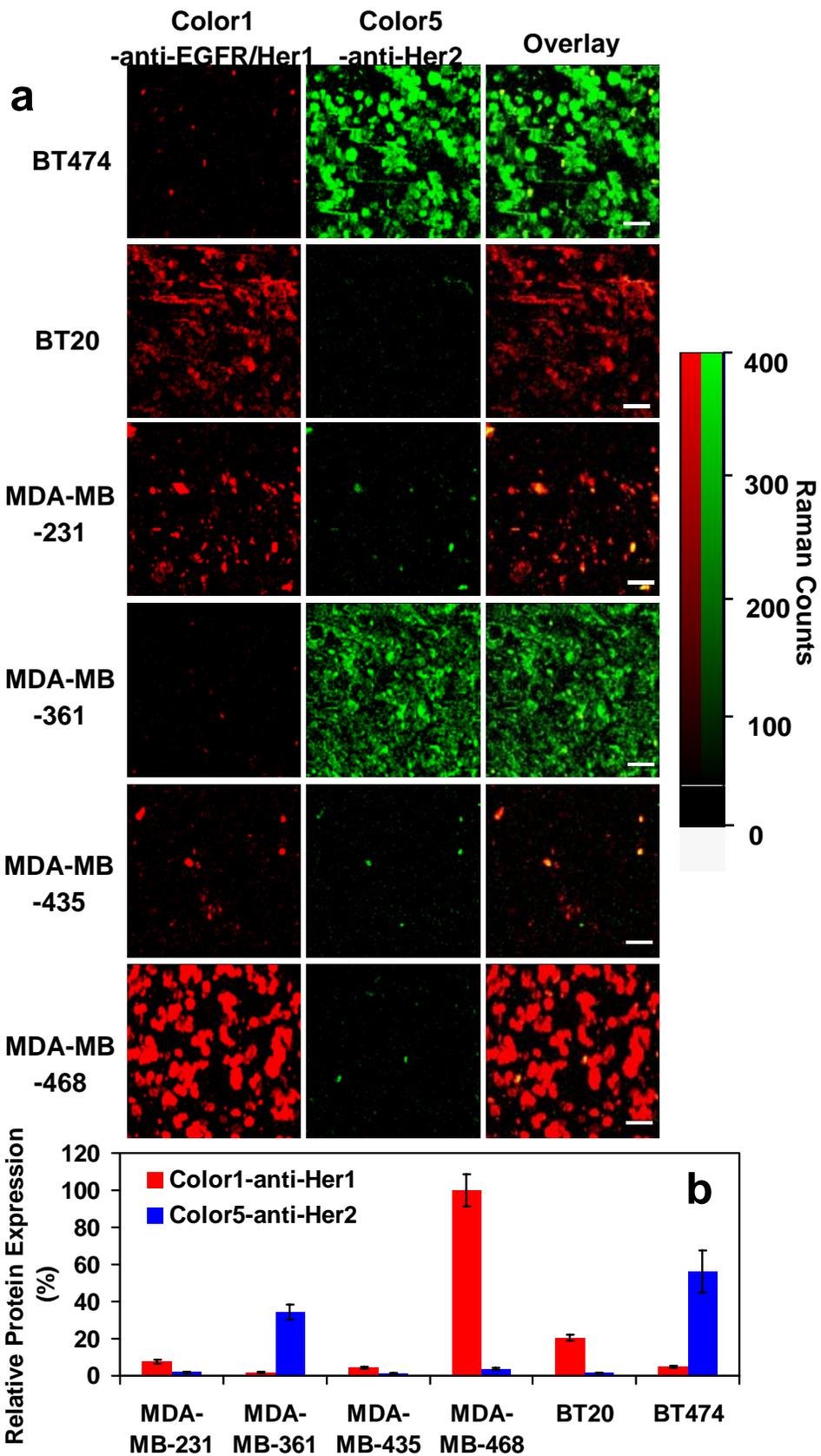

Figure 2

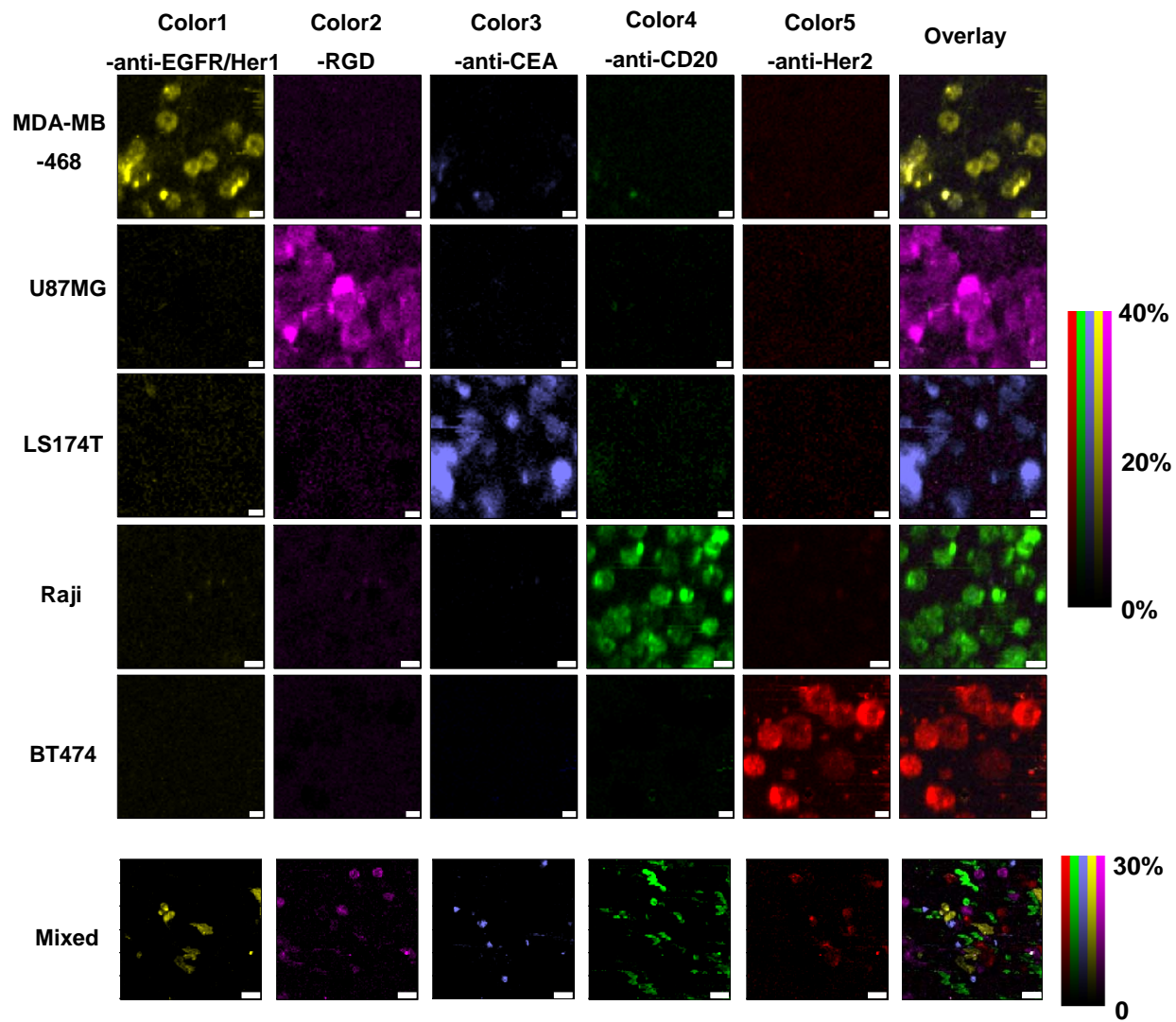

Figure 3

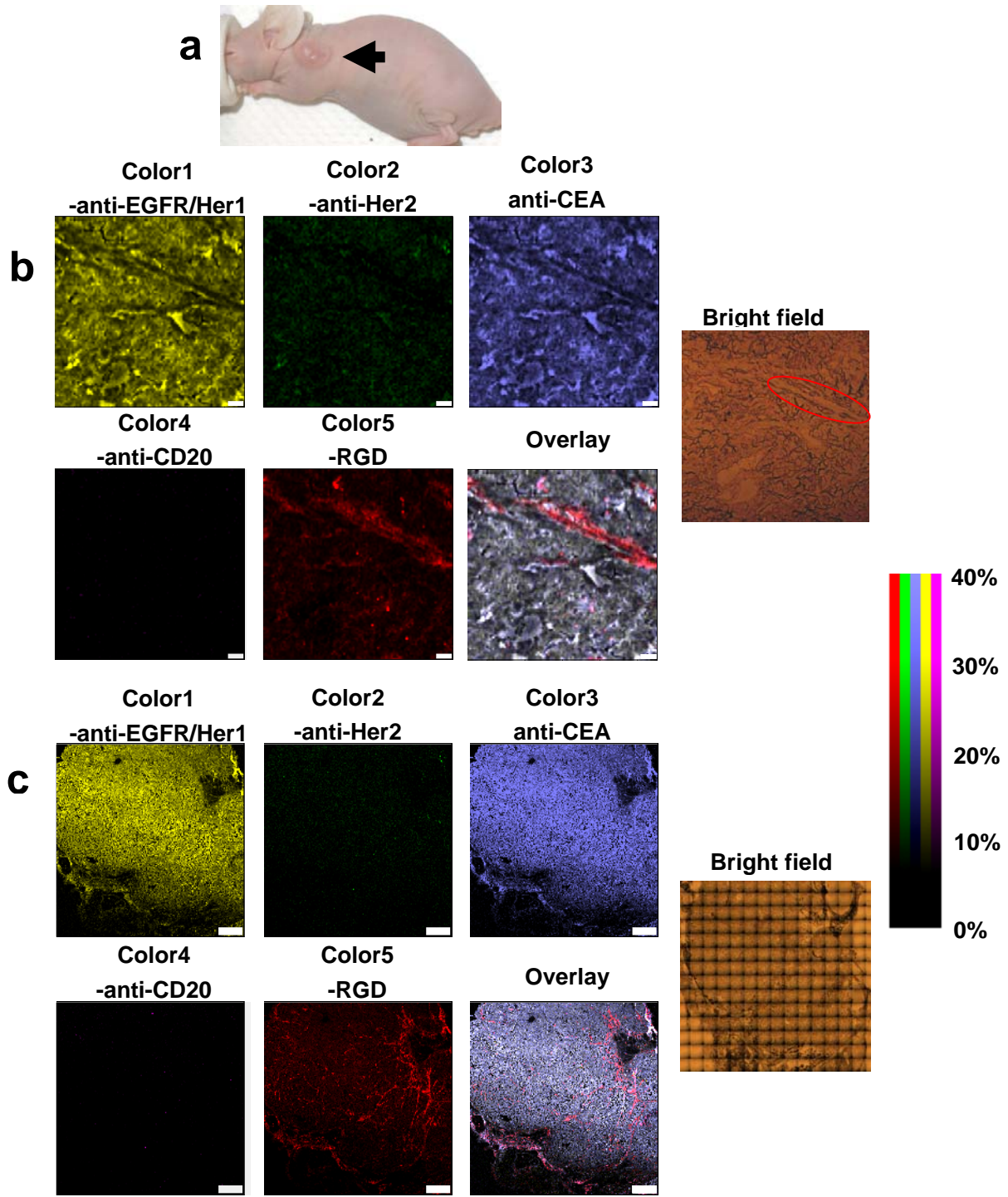

Figure 4

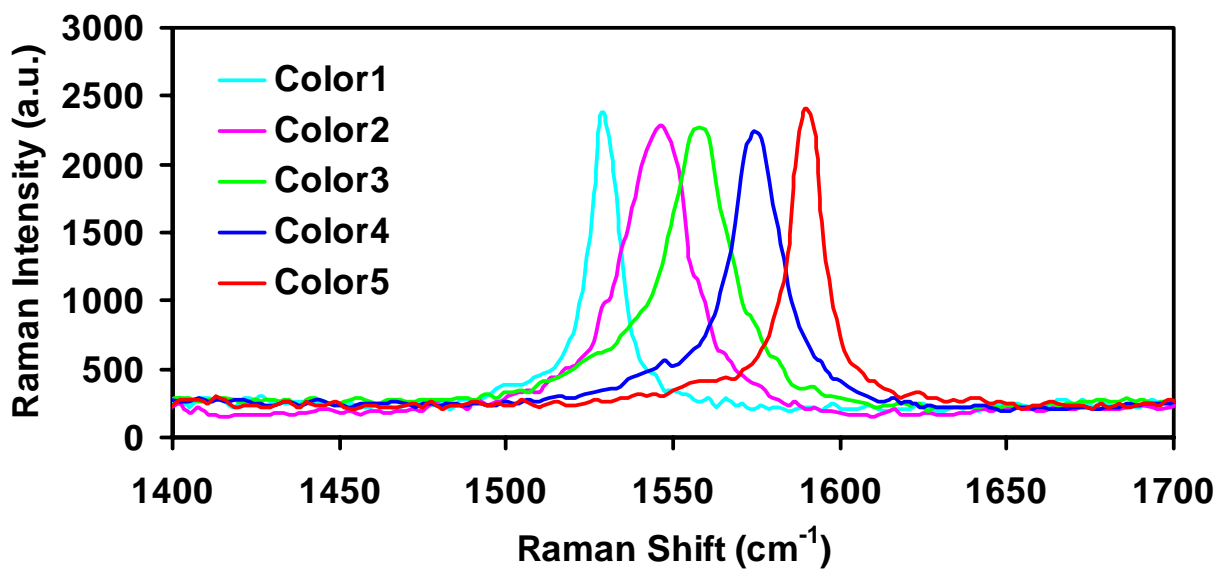

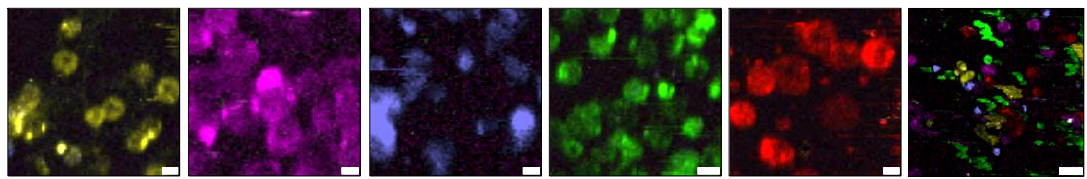

Isotopically-modified single-walled carbon nanotubes are synthesized and employed for biomedical Raman imaging. The five intense labels allow facile multiplexing in the near-infrared region, with minimal background interference.

Table of Contents Entry